\documentstyle[aps,preprint]{revtex}

\begin{document}
\author{M. Ruzzi}
\address{Depto. de F\'{\i}sica\\
Universidade Federal de Santa Catarina\\
88040 - 900 Florian\'{o}polis, S.C., Brazil}
\title{Schwinger, Pegg and Barnett and a relationship between angular and Cartesian
quantum descriptions. }
\maketitle

\begin{abstract}
From a development of an original idea due to Schwinger, it is shown that it
is possible to recover, from the quantum description of a degree of freedom
characterized by a finite number of states ({\it i.e}., without classical
counterpart) the usual canonical variables of position/momentum {\it and}
angle/angular momentum, relating, maybe surprisingly, the first as a limit
of the later.

\medskip\ 

\end{abstract}

\tighten

\section{Introduction}

Quantum mechanics has a lot of intriguing aspects. Definitely, one of those
aspects is the fact that there's still a handful of fundamental questions
about it, over which debate have not ceased after so many years. Among these
questions is the problem of the quantum phase, which a few years ago had an
important chapter (but not the final, it seems) in its history with the
approach due to Pegg and Barnett\cite{pegg}.

Within the broad grasp of the Pegg and Barnett formalism (PB), there is the
particular and important problem of one dimensional angular coordinates in
quantum mechanics. This specific problem is less problematic than the
question of the phase as a whole, but nevertheless it is also `solved' (or
re-solved) within the procedure of PB. In this article, I shall relate the
PB approach, in this particular context, to an idea presented by Schwinger,
and from this relation, although relatively simple, seems to emerge quite
interesting results.

Schwinger's original idea was to recover a usual Cartesian degree of freedom
(e.g., a degree of freedom endowed with a canonically related pair of
observables of position and linear momentum) from a degree of freedom
described by a finite set of states (that is, without classical counterpart)
through a limiting process. Here, I extend his discussion, showing that the
Cartesian degree of freedom can in fact be recovered by an infinite number
of limiting processes. The referred relationship comes from noting that a
limiting element of those infinitely many processes which work for the
Cartesian case reproduces exactly the Pegg-Barnett approach for the
angle/angular momentum case. So, in this sense, a circle would be the limit
of a line and not the opposite.

There's a conceptual bonus in the Schwinger procedure to obtain the quantum
description of a Cartesian degree of freedom. Schwinger's approach to finite
and discrete's degrees of freedom is, in its roots\cite{schi2}, by nature
laid over quantum mechanical concepts: quantum state, incompatible
observables and unitary transformations. Once that, starting from this, one
obtains the quantum description of degrees of freedom with classical
counterpart, it is then as explicit as it is possible that no quantization
of classical quantities must be necessarily involved in such descriptions.
The PB approach to angular coordinates can also be seen by the same
perspective and therefore shares this virtue. If one can see both
descriptions (Cartesian and angular) as different manifestations of a same
situation, then there might be room for new interpretations of the ultimate
physical meaning of such mathematical structures.

\section{The Schwinger unitary operators bases and the discrete genesis of
the canonical variables}

Long time ago Schwinger has noticed that one can obtain a complete basis in
operator space out of a pair of unitary operators $U$ and $V$, which act on
each other sets of $N$ eigenvectors as 
\begin{equation}
V^s|u_n\rangle =|u_{n-s}\rangle ,\qquad U^s|v_n\rangle =|v_{n+s}\rangle
,\qquad n=0,1,...N-1  \label{1}
\end{equation}
where a cyclic notation is understood, 
\begin{equation}
|u_k\rangle \equiv |u_{k(%
\mathop{\rm mod}
N)}\rangle \qquad \mid v_m\rangle \equiv \mid v_{(m\;%
\mathop{\rm mod}
N)}\rangle .  \label{2}
\end{equation}
The operators have the roots of unity as eigenvalues 
\begin{equation}
U|u_k\rangle =\exp \left[ \frac{2\pi i}Nk\right] |u_k\rangle ,\qquad
V|v_k\rangle =\exp \left[ \frac{2\pi i}Nk\right] |v_k\rangle ,  \label{3}
\end{equation}
and therefore 
\begin{equation}
U^N=V^N=\hat{1}.  \label{4}
\end{equation}
The pair also obeys Weyl algebra 
\begin{equation}
U^jV^l=\exp \left[ \frac{2\pi i}Njl\right] V^lU^j,  \label{5}
\end{equation}
and its eigenvectors are connected by a discrete Fourier transform 
\begin{equation}
\langle v_k|u_n\rangle =\frac 1{\sqrt{N}}\exp \left[ -\frac{2\pi i}N%
kn\right] ,  \label{6}
\end{equation}
which means that the two sets of states carry a maximum degree of
incompatibility. It must be kept clear that the this construction is
absolutely general, as Schwinger {\em obtains} all results above from the
mere existence of a complete family (with a finite number) of eigenstates of
a given abstract operator.

Schwinger has realized that the pair of operators $\{U,V\}$ could be used to
define a basis in {\it operator} space (as will be discussed in more detail
in a following work) and has also noticed that, if ones goes from this
discrete finite dimensional case to a usual continuous degree of freedom,
the ordinary position-momentum description is recovered.

To further extend Schwinger's original idea (which he concisely explored in
just a few lines), first it must be introduced a scaling factor 
\begin{equation}
\epsilon =\sqrt{\frac{2\pi }N},  \label{7}
\end{equation}
which will become infinitesimal as $N\rightarrow \infty $. Then, two
Hermitian operators $\{P,Q\},$(for simplicity, odd $N^{\prime }$s will be
considered, as even values only require only a little more care and a
heavier notation), 
\begin{equation}
P=\sum_{j=-\frac{N-1}2}^{\frac{N-1}2}j\epsilon ^\delta p_0|v_j\rangle
\langle v_j|\qquad Q=\sum_{j^{\prime }=-\frac{N-1}2}^{\frac{N-1}2}j^{\prime
}\epsilon ^{2-\delta }q_0|u_{j^{\prime }}\rangle \langle u_{j^{\prime }}|,
\label{29}
\end{equation}
constructed out of the projectors of the eigenstates of $U$ and $V.$ $\delta 
$ is a free parameter which might assume any value in the open interval $%
(0,2)$ (the original Schwinger discussion is equivalent to setting $\delta
=1 $). $\{p_0,q_0\}$ are real parameters that might carry units of momentum
and position, respectively, and $\epsilon ^\delta p_0$ and $\epsilon
^{2-\delta }q_0$ are the distance between successive eigenvalues of the $P$
and $Q$ operators. With the help of these, we can rewrite the Schwinger
operators as 
\begin{equation}
V=\exp \left[ \frac{i\epsilon ^{2-\delta }P}{p_0}\right] \qquad U=\exp
\left[ \frac{i\epsilon ^\delta Q}{q_0}\right] .  \label{28}
\end{equation}
Let also both eigenstate sets be relabeled as 
\begin{equation}
|v_j\rangle \equiv |p\rangle \qquad |u_{j^{\prime }}\rangle =|q\rangle
,\quad \text{with }q=q_0\epsilon ^{2-\delta }j^{\prime }\text{ and }%
p=p_0\epsilon ^\delta j.  \label{30}
\end{equation}
With that, 
\begin{equation}
P=\sum_{p=-\frac{N-1}2\epsilon ^\delta p_0}^{\frac{N-1}2\epsilon ^\delta
p_0}p|p\rangle \langle p|\qquad Q=\sum_{q=-\frac{N-1}2\epsilon ^{2-\delta
}q_0}^{\frac{N-1}2\epsilon ^{2-\delta }q_0}q|q\rangle \langle q|,
\label{30.5}
\end{equation}
and Eqs. (\ref{3}) now will read 
\begin{equation}
\exp \left[ \frac{ip^{\prime }Q}{p_0q_0}\right] \mid p\rangle =\mid
p+p^{\prime }\rangle \;\;\;  \label{31}
\end{equation}
and 
\begin{equation}
\exp \left[ \frac{iq^{\prime }P}{p_0q_0}\right] \mid q\rangle =\mid
q-q^{\prime }\rangle .  \label{32}
\end{equation}
if $\{p^{\prime },q^{\prime }\}$ are defined following the recipe of (\ref
{30}).

The equations above have a clear analogy with the usual relations between
position and momentum, apart from the fact that only discrete values of the
parameters are allowed and that the cyclic conditions (Eq.(\ref{2})) are
still holding.

The $N\rightarrow \infty $ limit can now be easily performed. For $\delta $
assuming any value in the open interval $(0,2)$, both Hermitian operators
defined on Eqs.(\ref{30.5}) will feature an unbounded and continuous
spectrum, as the limit leads them to\footnote{%
This limit has to be taken carefully, but it works as if the limiting value
of the discrete projector $|p\rangle \langle p|$ is, after the limit, $%
|p\rangle \langle p|dp.$ Just think on what happens to the resolution of
unity to be sure of that.}: 
\begin{equation}
P=\int_{-\infty }^\infty p|p\rangle \langle p|dp\qquad Q=\int_{-\infty
}^\infty q|q\rangle \langle q|dq  \label{30.75}
\end{equation}
and Eqs. (\ref{31},\ref{32}) now will be valid for any real numbers $%
\{p,q,p^{\prime },q^{\prime }\}.$ It must be observed that, in the way they
are obtained, the labels $\{p,q\}$ span the set of all rational numbers,
which is a proper subset of the set of real numbers. On the other hand,
every real number can be written as the limit of an infinite sequence of
rational numbers. Then the expression 
\begin{equation}
\exp \left[ \frac{i(p^{\prime }+p^{^{\prime \prime }}+p^{^{\prime \prime
\prime }}+...)Q}{p_0q_0}\right] \mid p\rangle =\mid p+p^{\prime
}+p^{^{\prime \prime }}+p^{^{\prime \prime \prime }}+...\rangle  \label{ref1}
\end{equation}
might converges to any real eigenvalue and its associated eigenvector. This
is enough to ensure that the hole usual Hilbert space of usual canonical
variables is recovered\footnote{%
The author would like to thank one of the anonymous referees for drawing
attention to this point.}\cite{Bohm}. Also, after the limit is performed the
cyclic condition becomes irrelevant, and the familiar relations are easily
recovered from their discrete counterparts

\begin{equation}
Q\mid q\rangle =q\mid q\rangle ,\quad \langle q^{^{\prime }}\mid q\rangle
=\delta \left( q^{^{\prime }}-q\right) ,\quad -\infty \leq q^{^{\prime
}},q\leq \infty  \label{33}
\end{equation}
\begin{equation}
P\mid p\rangle =p\mid p\rangle ,\quad \langle p^{^{\prime }}\mid p\rangle
=\delta (p^{^{\prime }}-p),\quad \langle p\mid q\rangle =\frac 1{\sqrt{2\pi
p_0q_0}}\exp \left( \frac{ipq}{p_0q_0}\right) .  \label{34}
\end{equation}
Therefore the results for a degree of freedom endowed with a usual
position-momentum canonical pair of variables are completely reproduced,
provided that the product of the parameters $p_0q_0$ is set to $\hbar $.

The $\epsilon ^{2-\delta }$ and $\epsilon ^\delta $ factors, roughly
speaking, control how `fast' (as $N$ increases) one will not be able to
identify the distance between labels of consecutive eigenvalues. The result
above is then rather peculiar, as it states that {\it how} you perform this
limit doesn't affect the final result. The usual canonical variables would
be recovered anyway.

But things can get different if you consider the extreme situation $\delta
=0 $ (or $\delta =2$, which is equivalent). In this case one of the
variables is not scaled at all and what follows is almost identical to the
Pegg-Barnett scheme (for simplicity, the reference angle is set to zero).
One would have 
\begin{equation}
V=\exp \left[ \frac{i\epsilon ^2M}{m_0}\right] \qquad U=\exp \left[ \frac{%
i\Theta }{\theta _0}\right]  \label{28}
\end{equation}
where 
\begin{equation}
M=\sum_{j=-\frac{N-1}2}^{\frac{N-1}2}jm_0|v_j\rangle \langle v_j|\qquad
\Theta =\sum_{j^{\prime }=-\frac{N-1}2}^{\frac{N-1}2}\epsilon ^2j^{\prime
}\theta _0|u_{j^{\prime }}\rangle \langle u_{j^{\prime }}|.  \label{29.5}
\end{equation}
If desired, the exponential of the angle operator might be used instead of
the operator itself, for the well known reasons given in \cite{carrut}. The
pair $\{m_0,\theta _0\}$ may carry different dimensional units. Let (again)
both eigenstates sets be relabeled as 
\begin{equation}
|v_j\rangle \equiv |m\rangle \qquad |u_{j^{\prime }}\rangle =|\theta \rangle
,\quad \text{with }\theta =\theta _0\epsilon ^2j^{\prime }\text{ and }m=m_0j.
\label{30b}
\end{equation}
In the $N\rightarrow \infty $ limit one would have 
\begin{equation}
M=\sum_{m=-\infty }^\infty m|m\rangle \langle m|\qquad \Theta =\int_{-\pi
}^\pi \theta |\theta \rangle \langle \theta |d\theta .  \label{30.5b}
\end{equation}

\begin{equation}
\Theta \mid \theta \rangle =\theta \mid \theta \rangle ,\quad \langle \theta
^{^{\prime }}\mid \theta \rangle =\delta \left( \theta ^{^{\prime }}-\theta
\right) ,\quad -\pi \leq \theta ^{^{\prime }},\theta \leq \pi  \label{33}
\end{equation}

\begin{equation}
M\mid m\rangle =m\mid m\rangle ,\quad \langle m^{^{\prime }}\mid m\rangle
=\delta _{m^{^{\prime }},m},\quad -\infty \leq m^{^{\prime }},m\leq \infty
\label{34}
\end{equation}
\begin{equation}
\langle \theta \mid m\rangle =\frac 1{\sqrt{2\pi m_0\theta _0}}\exp \left( 
\frac{i\theta m}{m_0\theta _0}\right) .  \label{34.5}
\end{equation}
The cyclic notation becomes meaningless to the $\mid m\rangle $ states in
the $N\rightarrow \infty $ limit, as this label gets unbounded. In the $\mid
\theta \rangle $ states, however, it takes naturally into account the
boundary conditions one good set of angle states must have, {\it i.e.,} 
\begin{equation}
\mid \theta \rangle \equiv \mid \theta (%
\mathop{\rm mod}
2\pi )\rangle ,  \label{35}
\end{equation}
and the action of the angle shift operator naturally obeys the boundary
condition. But it has to be stressed that (as in the Pegg-Barnett scheme),
the range of the variable $\theta $ is confined to $[0,2\pi )$ by {\it %
definition}, and ciclicyty modulo $2\pi $ is only matter of notation.
Therefore, and maybe surprisingly, the usual results for angle-angular
momentum variables are recovered from the same discrete root from which the
position-momentum results also emerged. Again, the product $m_0\theta _0$
must be set to $\hbar .$ $\theta _0$ is not expected to be a dimensional
unit but must be related to how one is measuring the angle.

\section{Conclusions}

The basic result here was to show that the two kinds of canonical variables
defined on degrees of freedom {\em with} classical counterpart can be
obtained from a description of a degree of freedom {\em without} classical
counterpart. In a pragmatic sense, one could say that the Pegg-Barnett
formalism for the angle/angular momentum case was seen as an extension of
the Schwinger approach to quantum Cartesian variables. In addition, the
discussion which led to those results have interesting aspects of its own.

One of those aspects is the role of the scaling factors in the limiting
process. In the first part of the discussion, where the parameter $\delta $
is free to vary in the open interval $(0,2)$, the initial discrete variables
are changed to a position/momentum like description, still discrete and with
contour conditions holding prior to effectively considering the limit. The
parameter $\delta $ controls the distance between successive eigenvalues of
the Hermitian operators $P$ and $Q$, and the greater the one, the smaller
the other, in such a way that their product is fixed. The infinite and
continuum limit of these variables is the position-linear momentum pair.
Schwinger had already stated that this would happen for $\delta =1$, and
what is surprising is that it happens to any value of $\delta $ in the open
interval $(0,2)$.

In the second part, we consider $\delta $ in one extreme of the interval
previously considered, $(\delta =0).$ Variables are now changed to an
angle/angular momentum like description. The limit to continuum in this case
only affects one of the variables (in the discrete/continuous sense) and the
angle/angular momentum operators and eigenstates are promptly recovered,
basically reproducing the PB scheme. The first interesting thing is that, in
this sense, angle/angular momentum variables are a limiting case of
Cartesian variables and not the opposite. One also sees that, for a finite
number of states, there is no fundamental distinction between angular or
Cartesian coordinates, or better, between the variables that will be
identified with angular or Cartesian coordinates {\em after} the limit is
taken, as representations (\ref{29.5}) and (\ref{30.5}) (prior to the $%
N\rightarrow \infty $ limit) can always be connected by a simple
transformation. The possibility of this transformation is only lost after
the limiting process.

As a parallel remark, there is nothing on the simple steps that led from
discrete to continuous variables that constrains the product of $p_0q_0$ to $%
\hbar .$ In fact, there's no (technical) reason for this product to have the
same value in both situations. We know from experience that this happens,
but it {\it could} be the case that $\hbar $ had a dependence on the number
of states allowed to the system (but fortunately it seems that is not).

In the sense above, one could say that it is not the geometry of a given
system that impose different quantum variables (in a quantization procedure
over a infinite line or over a ring), but, rather, that are the different
limiting cases of genuine discrete quantum descriptions that suit different
geometries. The author cannot refrain himself from remarking that even the
physical validity of this limit might be still put into discussion \cite
{leafc}.

After this work was finished the author became aware of reference \cite
{barker}, which discusses in great detail a similar limit to continuum, from
a mathematical point of view.

{\bf Acknowledgment:} The author credits Prof. F. F. de Souza Cruz for a
careful reading of the manuscript and valuable suggestions. With this work
the author would like to thank Prof. D. Galetti for years of fruitful
collaboration.

\end{document}